\documentclass[12pt,preprint]{emulateapj} 

\usepackage{lscape}
\usepackage{graphicx}
\usepackage{natbib}
\nocite{*}

\newcommand{\etal}{et~al.\/}

\newcommand{\oii}{\hbox{[O\,{\sc ii}}]}

\newcommand{\sii}{\hbox{[S\,{\sc ii}}]}

\newcommand{\lam}{$\lambda$}

\shortauthors{Werk \etal}
\shorttitle{NGC 2915}

\begin{document} 
\slugcomment{Accepted by ApJ: April 2010}

\title{The Metal-Enriched Outer Disk of NGC 2915}

\author{J. K.\ Werk\altaffilmark{1, 2},
M. E.\ Putman\altaffilmark{2},
G. R.\ Meurer\altaffilmark{3,4},
D. A.\ Thilker\altaffilmark{3},
R. J.\ Allen\altaffilmark{5},
J. Bland-Hawthorn\altaffilmark{6},
A. Kravtsov\altaffilmark{7},
K. Freeman\altaffilmark{8}}

\altaffiltext{1}{Department of Astronomy, University of Michigan, 500 Church St., 
		Ann Arbor, MI 48109, $jwerk@umich.edu$}
\altaffiltext{2}{Department of Astronomy, Columbia University, 550 West 120th Street, New York, NY 10027, USA}
\altaffiltext{3}{Department of Physics and Astronomy, The Johns Hopkins University, Baltimore, MD 21218-2686, USA}
\altaffiltext{4}{present address: ICRAR/The University of Western Australia, 35 Stirling Highway, Crawley, WA 6009, Australia}
\altaffiltext{5}{Space Telescope Science Institute, 3700 San Martin Drive, Baltimore, MD 21218, USA}
\altaffiltext{6}{Sydney Institute for Astronomy, School of Physics, University of Sydney, Australia}
\altaffiltext{7}{Kavli Institute for Cosmological Physics and Department of Astronomy and Astrophysics, The University of Chicago, 5640 S. Ellis Ave. Chicago, IL 60637}
\altaffiltext{8}{Research School of Astronomy and Astrophysics, Australian National University, Cotter Road, Weston Creek, ACT 2611, Australia}

\begin{abstract}
We present optical emission-line spectra for outlying HII regions in the extended neutral gas disk surrounding the blue compact dwarf galaxy NGC 2915. Using a combination of strong-line R23 and direct oxygen abundance measurements, we report a flat, possibly increasing,  metallicity gradient out to 1.2 times the Holmberg radius. We find the outer-disk of NGC 2915 to be enriched to a metallicity of 0.4 Z$_{\odot}$.  An analysis of the metal yields shows that the outer disk of NGC 2915 is overabundant for its gas fraction, while the central star-foming core is similarly under-abundant for its gas fraction. Star formation rates derived from very deep $\sim14$ ks GALEX FUV exposures indicate that the low-level of star formation observed at large radii is not sufficient to have produced the measured oxygen abundances at these galactocentric distances.  We consider 3 plausible mechanisms that may explain the metal-enriched outer gaseous disk of NGC 2915: radial redistribution of centrally generated metals, strong galactic winds with subsequent fallback, and galaxy accretion. Our results have implications for the physical origin of the mass-metallicity relation for gas-rich dwarf galaxies. 

\end{abstract}

\keywords{galaxies: abundances --- galaxies: dwarf --- galaxies: evolution --- galaxies: individual (NGC 2915) --- HII Regions --- ISM: HI}

\section{Introduction}
\label{intro}
The mass fraction of heavy elements in a galaxy is a fundamental physical property: it contains a record of the star formation history and therefore serves as an indicator of its evolutionary state. Numerous observational studies of gas metal abundances in a wide variety of galaxies, spanning a large range in mass and morphological types, have established a now well-known, yet poorly understood global relation in which the mean metallicity and stellar mass are directly correlated (e.g. Lequeux et al. 1979, Tremonti et al. 2004, Geha et al. 2009\nocite{lequeux79,tremonti04, geha09}).  The physical processes that govern this mass-metallicity relation are under debate, yet there is some consensus that low-mass dwarf galaxies hold the key to understanding this relation. Using a sample of 53,400 star-forming Sloan Digital Sky Survey galaxies, \cite{tremonti04} argue that metal loss via galactic winds is most likely to be responsible for the steep decline in the ``effective yield" (a parameterization of the metallicity divided by the galaxy total gas fraction) with galaxy baryonic mass. The physical basis for these outflows would be the cumulative effect of supernovae in disk OB associations which leads to low-mass galaxies with shallow gravitational potential wells selectively losing their metals \citep{maclow99,strickland04,brooks07}. Nonetheless, there are several other possible mechanisms for explaining the correlation between low metal yields and galaxy mass, including metal mixing in extended gaseous disks \citep{tassis08}, and a lower effective stellar upper mass limit to the initial mass function in dwarf galaxies \citep{koppen07, meurer09}. 


The as-yet open question of whether the extended neutral gas disks of dwarf galaxies are metal-enriched is quite relevant to this discussion, since each proposed model makes a prediction for the metal distributions in galaxies. For instance, in the SN-blowout scenario, metals tend to get ejected into the IGM for galaxies less massive than 10$^{9}$ M$_{\odot}$ \citep{maclow99} or 10$^{11}$  M$_{\odot}$ \citep{strickland04}, while in the metal-mixing scenario, the central regions and outermost regions of the dwarf galaxy should have the same metal abundance. Therefore, abundance gradients (or a lack thereof) in the extended gaseous disks of dwarf galaxies may help to untangle the physical origin of the mass-metallicity relation. 


 
\begin{figure*}[t]
\epsscale{0.8}
\plotone{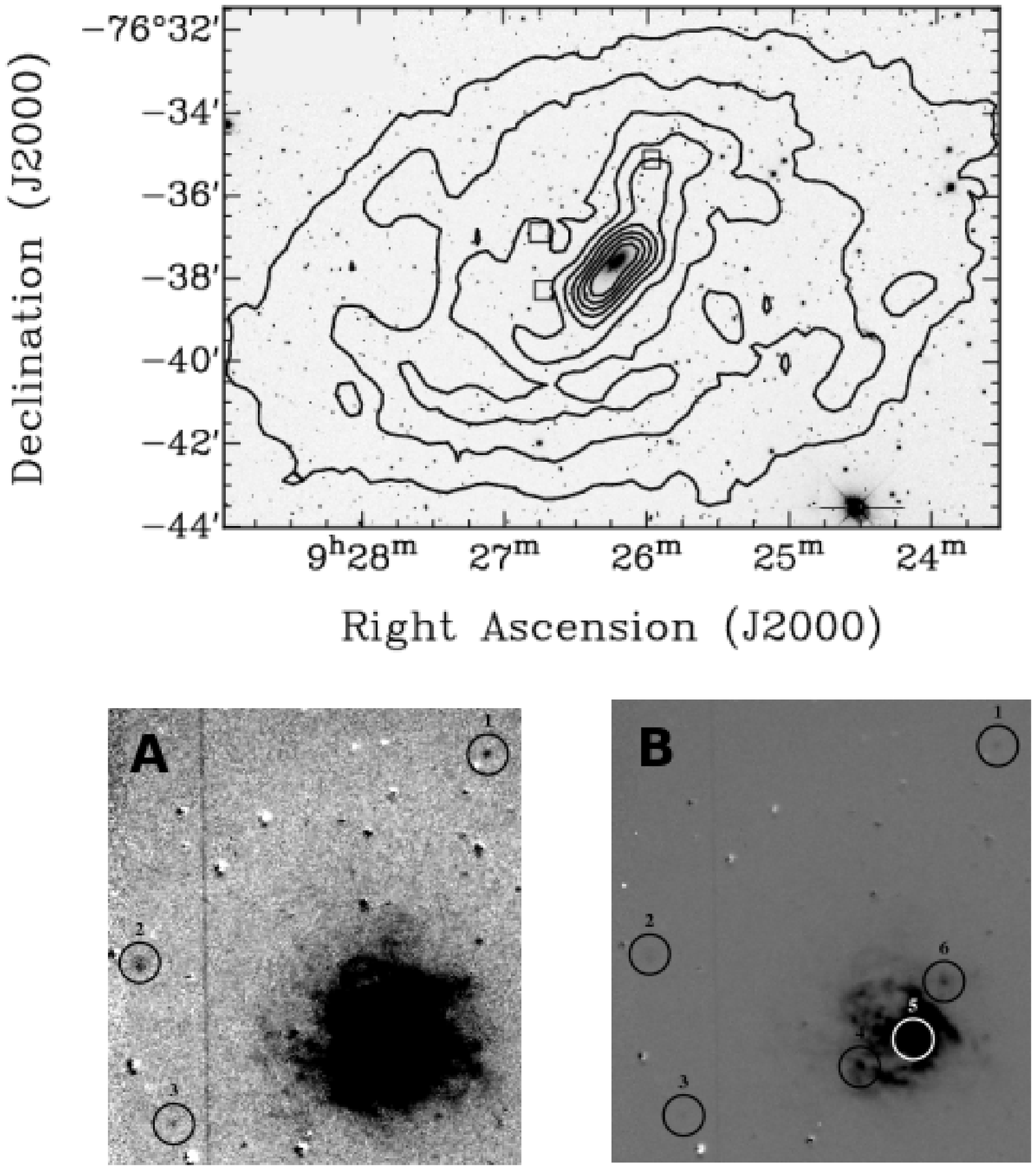}
\caption{Top: A CTIO 1.5-m broadband optical V-Band image of NGC 2915 overlaid with ATCA HI contours. The HI contour levels are 5, 15, 25, 35, 45, 55, 65, 75, 85, and 95 percent of the peak HI column density, 1.71 $\times$10$^{21}$ cm$^{-2}$.  The outer gaseous disk of NGC 2915 mapped with the ATCA appears to extend 5 times beyond the optical component of the galaxy which has a 2.3 kpc extent. On the image, we have boxed the 3 outlying HII regions included in this study, and contained in the lower zoomed-in panels. Bottom: H$\alpha$ continuum-subtracted images of the core of NGC 2915 shown in two stretches: (A) to highlight the outer HII regions included in this study, and (B) to highlight the inner HII regions included in this study. }
\end{figure*}

 NGC 2915 is one of the most extreme examples of a blue compact dwarf galaxy with an extended gaseous disk (Meurer et al. 1996; hereafter M96\nocite{meurer96}).  This nearby (4.1 Mpc, Meurer et al. 2003 \nocite{meurer03}) dwarf galaxy has an HI disk that extends 5 times beyond the optical stellar component (12 kpc for the gas; 2.3 kpc for the stars; see Figure 1, and Table 1 for a list of its full properties). 
 \begin{deluxetable}{llr}
\tablecaption{NGC 2915 Optical and HI Properties \label{props}}
\tabletypesize{\scriptsize}
 \tablewidth{0pt}
  \tablehead{\colhead{Property} &
  \colhead{} &
  \colhead{Value}}
\startdata
        RA (J2000)& &09:26:11.5\\
        dec (J2000)&&-76:37:35\\
        distance&&4.1$\pm$0.3 Mpc\\
        Holmberg Radius&&2.3 kpc\\
        M$_{\star}$&&3.2$\times$10$^{8}$ M$_{\odot}$\\
        M$_{gas}$&& 7.4$\times$10$^{8}$ M$_{\odot}$\\
        E(B-V)&&0.275$\pm$0.04\\
        M$_{V}$& &-16.42\\
        M$_{dyn}$/L & & 62* (solar)\\
        R$_{HI}$&&11.9 kpc\\
        
\enddata
\tablecomments{Properties from M03 and M96, in some cases updated for a revised distance measurement. (*) New HI data from \cite{elson10} indicates that M$_{dyn}$/L may in fact be as high as 140 (solar units). Reddening comes from \cite{schlegel98}.}
\end{deluxetable}
\begin{deluxetable*}{lccccc}[b]
\tablecaption{NGC 2915 HII Regions \label{list}}
\tabletypesize{\scriptsize}
 \tablewidth{0pt}
  \tablehead{\colhead{HII Region} &
  \colhead{RA} &
  \colhead{dec} &
  \colhead{R/R$_{Ho}$} &
  \colhead{Log L$_{H\alpha}$} &
  \colhead{notes} \\
   \colhead{} &
    \colhead{J2000} &
     \colhead{J2000} &
      \colhead{} &
       \colhead{(ergs s$^{-1}$)} &
        \colhead{}}
        \startdata
        1 & 09:26:00.3 & -76:35:25.0 & 1.2 & 36.2 & low ionization parameter; [OIII]$\lambda$4959 not detected \\
        2 & 09:26:45.9 & -76:37:01.1 & 1.1 & 36.5 &  diffuse, outer HII region; [OII]$\lambda\lambda$3727 detected at 3$\sigma$\\
        3 & 09:26:41.7 & -76:38:12.9 & 1.0 & 35.8  & only H$\alpha$ and [OIII]$\lambda$5007\\
        4 & 09:26:19.8 &  -76:38:02.0 & 0.34 & 38.4 & slit length = 30\arcsec, on edge of central star formation in NGC 2915\\
        5 & 09:26:11.1 & -76:37:39.0 & 0.04 & 39.2 & central, luminous HII region; [OIII]$\lambda$4363 detected\\
        6 & 09:26:07.3 & -76:37:12.9 & 0.23 & 37.8 &  bright HII region within galaxy\\
  \enddata
\tablecomments{Values of the total H$\alpha$ luminosity are derived from the TTF image, and are corrected for internal extinction derived from \cite{schlegel98}. Errors in the H$\alpha$ luminosity are on the order of 10\%\, dominated by the flux calibration errors. }
\end{deluxetable*}

 Its total baryonic mass (gas plus stars; $\sim10^{9}$ M$_{\odot}$) puts it on the high-mass end of the spectrum of dwarf galaxies, and its total dynamical mass gives it one of the highest-known mass-to-light ratios for a gas-rich galaxy (M96).  Recent H$\alpha$ images of NGC 2915 have revealed several small pockets of star formation embedded in its extended gaseous disk at projected radii of $\sim3$kpc that otherwise contains few stars (see Figure 1). In addition, new, very deep (t$_{exp}$ = $\sim14$ ks) images from the {\it{Galaxy Evolution Explorer}} Satellite ({\it{GALEX}}) show very faint extended-UV (XUV) emission near the location of these HII regions.  Because of their strong emission-line spectra, HII regions are relatively straightforward probes of the oxygen abundance of surrounding gas. In the most luminous cases, a direct measurement of the oxygen abundance can be made using the temperature sensitive, faint [OIII]$\lambda4363$ line, and in other cases, calibrated relations between strong lines (e.g. R23; Pagel et al. 1979; McGaugh 1991, hereafter M91 \nocite{pagel79,mcgaugh91}) can provide reasonable estimates of  the oxygen abundance. Dwarf galaxies have been observed to have spatially homogeneous oxygen abundances, although the most recent studies have so far  been limited to well-within the Holmberg radius \citep{lee06, croxall09}. By comparison, we report the oxygen abundances for HII regions out to 1.2 times the Holmberg radius (the isophotal radius at which the surface brightness is below 26.5 in the B-band) and discuss the implications of our results.

\section{Observations and Data Reduction}
\subsection{Imaging}

   We selected HII regions in and around NGC 2915 from H$\alpha$ images taken in February 1998 on the 3.9-m Anglo-Australian Telescope (AAT) with the Taurus Tunable Filter (TTF; Bland-Hawthorn \& Jones 1998 \nocite{blandhawthorn98}).  The images consist of a total of four 600$s$ exposures using an 11\AA\ filter centered on H$\alpha$ and two 90$s$ exposures in the  11 \AA\ continuum-band. These images were reduced using standard techniques for TTF images, aligned, and combined to make a single image spanning $7.2'$ (8.5 kpc) with $2.3''$ (46 pc) seeing. The data were flux calibrated to $\sim 10$\%\ accuracy using observations of two spectro-photometric standards.  Figure 1 shows both a broadband optical image of NGC 2915 overlaid with HI contours (M96) and two continuum-subtracted H$\alpha$ images of the galaxy that show the six HII regions presented in this study.  The broadband optical data (V, R, I-bands) were obtained in February 2003 on the CTIO 1.5-m telescope. The I-band data used for calculations in Section 4 are the sum of 6 exposures totaling 720s. Table 2 presents the locations, H$\alpha$ luminosities, and a brief description of the spectral properties of the HII regions.  
\begin{figure*}[h!]
\begin{center}
\vspace{-0.01in}
\plotone{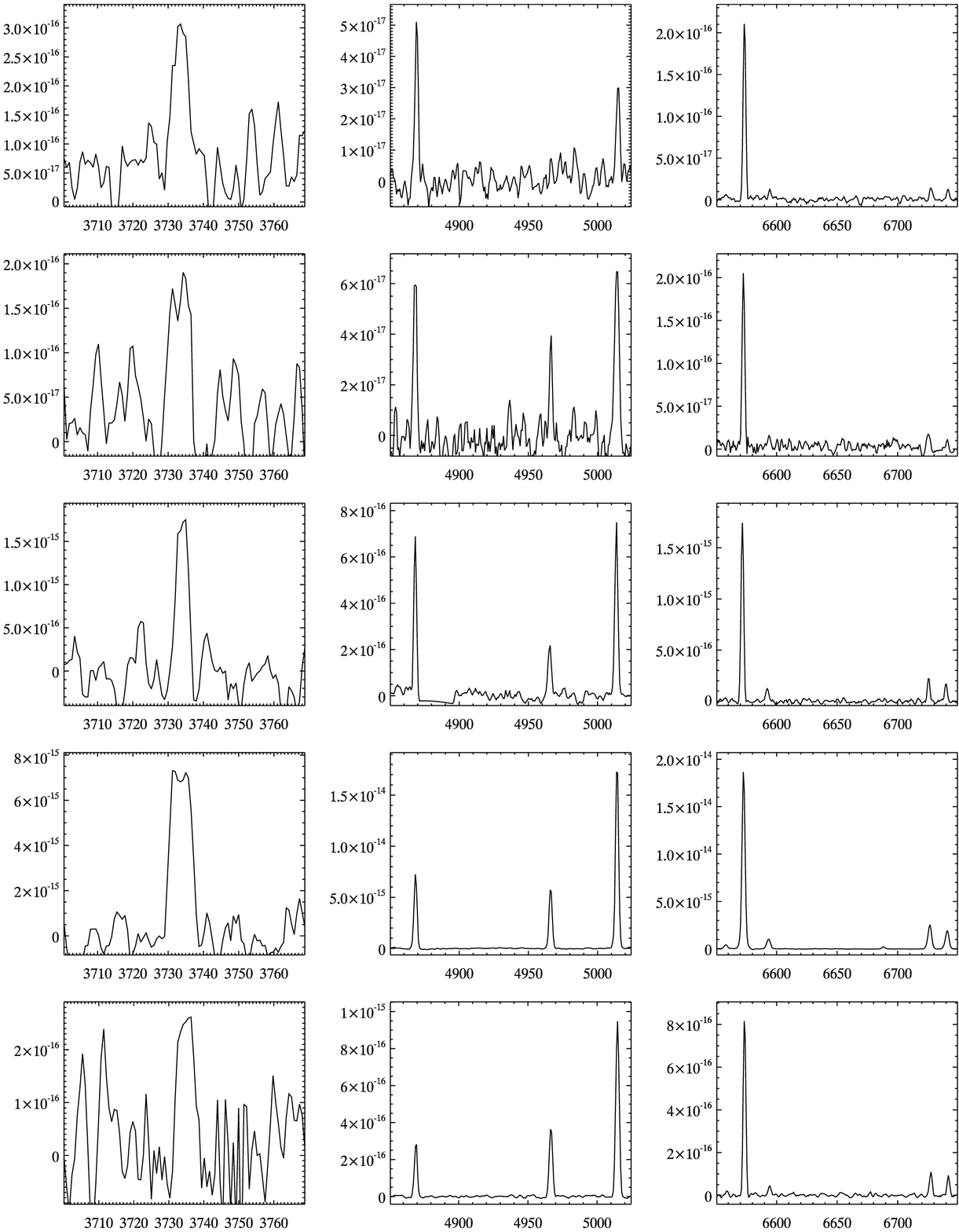}
\end{center}
\caption{\label{spectra}:  The multislit spectra reveal the essential emission lines for abundance determination. Left:  [OII] $\lambda\lambda$3727, Middle: H$\beta$, [OIII] $\lambda$4959 and $\lambda$5007, and Right: H$\alpha$, [NII] $\lambda$6583 and [SII]$\lambda\lambda$6717,6731. The spectra are arranged in descending order: outer HII Region 1 at the top, followed by outer HII Region 2, then central regions 4, 5, and 6 (bottom). Flux units are ergs s$^{-1}$ cm$^{-2}$ $\AA^{-1}$, and wavelength is in Angstroms. These one-dimensional calibrated spectra have been continuum-subtracted (inner regions), dereddened, and 3 $\times$ 3 boxcar-smoothed. }
\end{figure*}

The UV observations of NGC 2915 were obtained as part of the GALEX Deep Galaxy Survey (Thilker et al. in prep).  FUV and NUV images were recorded over the course of 12 independent visits and subsequently co-added using the GALEX pipeline (release ops$-$v6\_2\_0). Total exposure time is 13.8 ks in both bands.  The angular resolution of GALEX is 4.2"[5.3"] FWHM in FUV[NUV].  More details regarding the on-orbit performance of GALEX can be found in Morrissey et al. (2007). Foreground stars were manually masked within our region of interest prior to performing the UV photometry.

\subsection{Spectroscopy}

Multi-slit spectroscopy (0.7\arcsec~slitlets, 10\arcsec~long in most cases) was carried out on NGC 2915 over two clear nights in January 2008 with the Imamori Magellan Areal Camera and Spectrograph (IMACS) on the Baade 6.5-m telescope at Las Campanas Observatory. IMACS, in combination with the f/4 camera, provided a 15.4\arcmin~FOV that is well matched to the size of the field containing the extended HI envelope of NGC 2915 (R$_{HI}$ = 10 arcmin; M96). In order to detect and resolve both  \oii \lam\lam 3727 and \sii \lam\lam 6717,6731, we used  two 600 l/mm gratings, one tuned to the blue portion of the optical spectrum, and the other tuned to the redder side. The red side spectra cover 5430 \AA~ -- 8590 \AA~, while the blue side spectra cover 3360 \AA~-- 6470 \AA, both with a dispersion of 0.382 \AA~ per pixel.  The blue-red overlap coverage contains the faint HeI \lam5876 and [OI] \lam6300 emission lines, allowing us to examine the agreement of the red and blue flux calibration for the brightest central HII region in our sample. For both of the emission lines, the line flux agreement is excellent and well-within the typical uncertainty of $\sim7\%$ (see below; [OI], red (blue), is 2.97$\pm0.15$ (3.05$\pm0.17$)$\times$10$^{-15}$ ergs s$^{-1}$ cm$^{-2}$ \AA$^{-1}$ and HeI, red (blue), is 2.23$\pm0.12$ (2.24$\pm0.14$) $\times$10$^{-15} $ergs s$^{-1}$ cm$^{-2}$ \AA$^{-1}$ ). Since the HII regions are very faint, we required numerous long exposures: 7 $\times$ 2000 seconds in the red, and 7 $\times$ 3000 seconds plus 10 $\times$ 2000 seconds in the blue.
    
For the reduction of our multi-slit data, we used the COSMOS package, a set of programs developed by Carnegie Observatories and based on a precise optical model of IMACS. The COSMOS process is one of alignment, production of an accurate spectral map, basic reduction steps (i.e. bias subtraction, flat-field corrections), night sky line subtraction, and spectral summing and extraction. Arc lamps taken before and after each exposure ensure an accurate wavelength calibration. We performed flux-calibration on 1-d extracted, summed spectra using two standard stars, LTT 1788 and LTT 4816, both observed multiple times over the course of the observing run. The short length of the mask slitlets resulted in some emission lines occupying the full-length of the slit, therefore complicating night sky line subtraction. A number of strategically-placed blank slitlets on the field of NGC 2915 allowed us to properly subtract the night sky by fitting and subtracting profiles (IRAF task fitprofs) for  individual sky lines near emission-lines arising in NGC 2915. Finally, for the three HII regions more centrally-located, we corrected for absorption from underlying stellar populations by subtracting gaussian fits to the Balmer-line absorption.

\section{Analysis}
\label{anal}

For two of the faint, outer HII regions, and all three of the more centrally located HII regions, we were able to measure the strong recombination emission lines of the Balmer series (H$\alpha$ and H$\beta$), and forbidden metal lines of [OII] $\lambda\lambda$3727, [OIII] $\lambda$4959 and $\lambda$5007, [NII] \lam6583 and [SII]\lam6717 and \lam6731. In the three central HII regions, we detect H$\alpha$, H$\beta$, H$\gamma$ and H$\delta$ (only region5), in addition to several fainter lines of helium, neon, and argon. For example, in the spectrum of region 5, we detect fainter lines of helium (HeI \lam4026, \lam4472, \lam4922, \lam5876, \lam6678, \lam7065, and HeII \lam4687), neon ([NeIII] \lam3869, \lam3970), and argon ([ArIII] \lam 7135). Also for Region 5, we detect the very faint, temperature-sensitive [OIII] $\lambda$4363 auroral emission line which provides a direct oxygen abundance measure (see below). We detect only H$\alpha$ and [OIII] $\lambda$5007 in the spectrum of region 3, and therefore exclude this region from our subsequent oxygen abundance analysis, as these two emission lines are not sufficient for our analysis. Final, reduced, one dimensional spectra for the 5 HII regions included in this study are shown in Figure 2. The H$\beta$ emission line for Region 4 falls near, but luckily completely outside, one of the IMACS chip gaps (seen as a near straight line in the continuum of the center-most panel).

We apply a correction for interstellar reddening to all line measurements from the observed H$\alpha$ to H$\beta$ ratio for case B recombination where H$\alpha$/H$\beta$ = 2.86 at an effective temperature of 10,000 K and electron density of 100 cm$^{-3}$ \citep{hummer87}. We use a reddening function normalized at H$\beta$ from the Galactic reddening law of \cite{ccm} assuming R$_{v}$ = A$_{v}$/E(B$-$V) = 3.1. Ratios of H$\gamma$ to H$\beta$ were additionally examined in the three bright central HII regions (4, 5, and 6), and were found to give lower values of E(B-V): 0.14, 0.10, and 0.21 compared to 0.53, 0.35, and 0.26, respectively. Since we are directly comparing the measurements from the faintest HII regions with those of the more luminous central HII regions, we wished to use the same emission-line ratios for the analysis of the 5 HII regions included in this work. Therefore, in order to be consistent, we use only the H$\alpha$/H$\beta$ Balmer decrement to correct for extinction, while acknowledging that the error in the reddening could be up to 0.2 magnitudes. Nonetheless, the reddening has very little impact on the calculated strong-line and direct abundances presented below.  Table 3 presents the reddening-corrected strong line measurements for the 5 HII regions in this study. All listed line fluxes are relative to  H$\beta$ = 100. The errors given in the table account for  the RMS of the line measurements, the flux calibration uncertainty, read noise, sky noise, and flat-fielding errors. 

\subsection{Strong Line Oxygen Abundances}
We obtain the nebular oxygen abundances for four of the five regions presented in this study using the strong line R23 method originally presented by \cite{pagel79}, according to the calibration of M91.  R23 is defined as log [([OII] $\lambda\lambda3727$ + [OIII] $\lambda4959$ + [OIII] $\lambda5007$)/H$\beta$]. We choose M91 over the many other calibrations of the R23 relation based mainly on the analysis presented in \cite{vanzee06}, which concludes that the photoionization grid of M91 is more physically-accurate than that of \cite{pilyugin00}. The drawbacks of the R23 method include a well known degeneracy and turnover at $\sim0.3$Z$_{\odot}$ and systematic errors due to age effects and stellar distributions. These drawbacks have been discussed extensively in the scientific literature, and are not repeated here \citep{kewley08, ercolano07}. We break the degeneracy using the [NII]/[OII] line ratio for our HII regions. Since all of the HII regions have log [NII]/[OII] $<$ -1.0, all are assumed to lie on the lower metallicity branch of the R23 relation (M91). Errors in the R23 calibration are on the order of 0.2 dex in the turnover region (M91, Ercolano et al. 2007 \nocite{ercolano07}), but tend to be higher in regions that have low ionization parameters \citep{vanzee06}. Varying the reddening between 0.0 and 0.6 magnitudes randomly for all the HII regions results in values of 12 + Log(O/H) that differ from those quoted values by at most 0.1 dex. 

\subsection{Direct Oxygen Abundance for Region 5}
Since we have detected [OIII] $\lambda$4363 in the spectrum of Region 5, we are able to make a direct measurement of its oxygen abundance. The line flux of the [OIII] \lam4363 emission line is 7.8$\pm$0.54 $\times$ 10$^{-16}$ ergs s$^{-1}$ cm$^{-2}$ \AA$^{-1}$. We use the IRAF package NEBULAR, based on the FIVEL program \citep{fivel,shaw95}. We assume a 2-zone ionization model (high and low) in which the temperature of the O$^{++}$ zone is derived from the line strengths of the [OIII] emission lines \citep{osterbrock89}, and the effective temperature, T$_{e}$,  in the O$^{+}$ zone is given by T$_{e}$(O$^{+}$) = 2[(T$_{e}$(O$^{++}$))$^{-1}$ + 0.8]$^{-1}$  where T$_{e}$ is in units of 10$^{4}$ K \citep{stasinska90,pagel92}. We use the line ratio [SII] $\lambda6717$/$\lambda6731$ to determine the electron density, which is approximately 100 cm$^{-3}$. We obtain temperatures for the low and high ionization zones of 12900 K and 13400 K, respectively, which results in an intrinsic ratio H$\alpha$/H$\beta$ = 2.80 \citep{hummer87} for Region 5. Based on these values (which give emissivity coefficients) and the line strengths for Region 5, the IRAF package ABUND provides ionic abundances for oxygen from which we then compute the total oxygen abundance. Again, we note that the reddening has very little impact on the calculated oxygen abundance. 
 \begin{deluxetable*}{l c c c cc c c c c c  }
  \tablecaption{Reddening-Corrected Line Measurements \label{lines}}
  \tabletypesize{\scriptsize}
  \tablewidth{0pt}
  \tablehead{\colhead{HII} &
           \colhead{[OII]} &
          \colhead{[OIII]} &
           \colhead{[NII]} &
             \colhead{[SII]} &
             \colhead{[SII]} &
             \colhead{Log R23} &
             \colhead {Log O32} &
              \colhead{$f_{H\beta}$} &
             \colhead{E (B$-$V)} &
             \colhead{12 + Log(O/H)}  \\
             \colhead{reg.} &
             \colhead{3727} &
             \colhead{5007} &
             \colhead{6583} &
             \colhead{6717} &
             \colhead{6731} &
             \colhead{} &
             \colhead{} &
             \colhead{(ergs s$^{-1}$ cm$^{-2}$)} &
             \colhead{(mag)} &
             \colhead{M91 calibration}}
            
\startdata
1 & 480$\pm$55 & 40$\pm$3.9 & 18$\pm$2.8 & 34$\pm$3.4 & 25$\pm$2.7 & 0.721$\pm$0.06 & -0.939$\pm$0.07&(1.98$\pm$0.16)e-16 & 0.450 & 8.26$\pm$0.25\\
2 & 500$\pm$55 & 130$\pm$10 & 16$\pm$2.9 & 29$\pm$2.7 & 18$\pm$1.9 &0.830$\pm$0.07 &-0.453$\pm$0.04&(2.15$\pm$0.18)e-16 & 0.309  & 8.26$\pm$0.25\\
4 & 430$\pm$36 & 120$\pm$9.3 & 24$\pm$1.9 & 34$\pm$3.1 & 28$\pm$2.3 & 0.765$\pm$0.04&-0.403$\pm$0.03&(1.68$\pm$0.13)e-15 & 0.533 & 8.14$\pm$0.20\\
5 & 240$\pm$18 & 250$\pm$19 & 18$\pm$1.1 & 41$\pm$1.6 & 31$\pm$1.4 & 0.757$\pm$0.05&0.161$\pm$0.02& (2.11$\pm$0.16)e-14 & 0.350 & 7.94$\pm$0.05* \\
6 & 161$\pm$15 & 340$\pm$26 & 15$\pm$1.7 & 33$\pm$2.9 & 27$\pm$2.4 & 0.792$\pm$0.05&0.456$\pm$0.05&(7.86$\pm$0.59)e-16 & 0.258  & 7.88$\pm$0.20\\

  \enddata
\tablecomments{All line strengths are given in terms of H$\beta$=100. Reddening correction for case B recombination with T = 10000K and ne = 100 cm$^{-3}$, where H$\alpha$/H$\beta$ = 2.86. E(B-V) values compare well with the Schlegel value of 0.275, but may have errors as high as 0.2 magnitudes (see Section 3). These large errors in E(B-V) contribute minimally to the error in the strong-line abundances. Region numbers match those in Figure 1. In the case of region 5, the central-most luminous HII region, we were able to determine the oxygen abundance via the direct method outlined in Section \ref{anal}. Typical errors in the R23 method as defined by M91 are near 0.2 dex at the break in the R23 relation (at 12 + Log(O/H) $\sim8.3$), and are thus the largest contributors to the error in the oxygen abundance. The error in the R23 calibration increases for nebulae with very low ionization parameters. In this table, we have estimated the error of  12 + Log(O/H) to be between 0.20 and 0.25, except for region 5, which has a more accurate direct determination of its oxygen abundance. *A direct abundance value, with associated error. The M91 value for this region matches the direct abundance almost exactly, at 12 + Log(O/H) = 7.93. }
\end{deluxetable*}

  \section{Results}
 \begin{figure}[t]
\epsscale{1.2}
\plotone{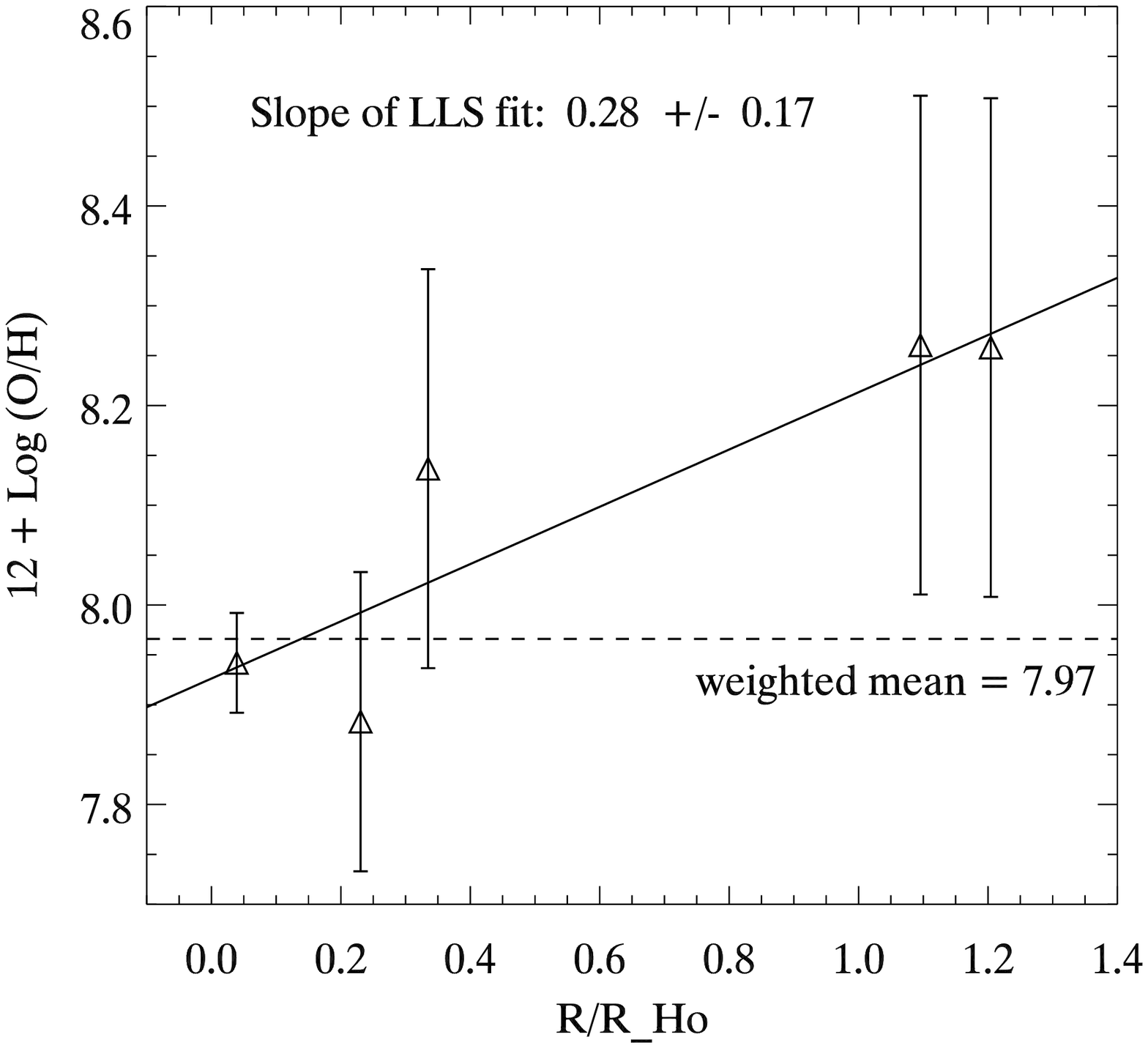}
\figcaption{\label{radgrad} The oxygen abundance, 12 + Log (O/H), versus the galactocentric projected distances (in R/R$_{Ho}$) for the 5 HII regions for which we could measure the necessary strong lines (5, 4, 6, 2, 1, in order of R/R$_{Ho}$). The errors are largest for the two outermost HII regions because they are both located on the turnover of the R23 relation, and because they have very low ionization parameters, which makes the R23 abundance determination more uncertain \citep{vanzee06}. The dashed line represents the mean 12 + Log (O/H) value, weighted by 1/$\sigma^{2}$,  for the 5 HII regions. The solid line represents a linear-least-squares fit to the data, the slope of which is consistent, within 2$\sigma$, with a flat radial oxygen abundance gradient in NGC 2915 out to 1.2 times its Holmberg radius. }
\end{figure}

  Table 3 lists the derived strong-line R23 (M91) abundances for Regions 1, 2, 4, and 6, and the direct oxygen abundance obtained for Region 5. We compare the direct value for Region 5 with its R23 strong-line value, and find they are nearly identical: 12 + Log (O/H) direct = 7.94 $\pm$ 0.05 and 12 + Log (O/H) strong-line = 7.93 $\pm$ 0.20. Assuming 12 + Log (O/H)$_{\odot}$ = 8.66 \citep{asplund}, we find that the 5 HII regions have oxygen abundances that range from 0.2 to 0.4 Z$_{\odot}$. These values are roughly consistent with several other strong-line abundance methods, including the N2 index obtained using  the [NII] \lam6583/H$\alpha$ ratio \citep{pettini04} which gives an oxygen abundance of $\sim0.4$ Z$_{\odot}$ for all 5 HII regions. The inverse variance weighted mean of the 5 oxygen abundance measurements is 7.97 $\pm$ 0.05. We present the oxygen abundances as a function of galactocentric distance in Figure 3.  The linear-least-squares fit to these data, including their errors (performed with the IDL routine {\it fitexy}), shows that NGC 2915 has an increasing radial oxygen abundance gradient out to 1.2 Holmberg radii. However, due to the large errors inherent in the strong-line abundances, this slope (0.28 $\pm$ 0.17), within 2$\sigma$, is also consistent with a flat abundance gradient in NGC 2915. Nonetheless, the scatter of the five points about the best-fit line is much smaller than the error bars. For this reason,  we consider that we may be overestimating the errors, and perform another inverse variance weighted linear least squares fit on the data (assuming the error values are now {\it relative}).  The resultant slope is unchanged, although $\sigma_{slope}$ is decreased from 0.17 to 0.07. This smaller $\sigma_{slope}$ may be stronger evidence for an increasing abundance gradient, but only if we are overestimating the errors on the R23 oxygen abundances (as the small scatter about the best-fit line may indicate). We consider it quite unlikely, based on arguments by \cite{ercolano07} and others, that the errors in the strong line abundances are significantly less than 0.2 - 0.3 dex. 
  
 Figure 4 plots effective yield versus total baryonic mass for NGC 2915 (filled circle), assuming its total oxygen abundance is the weighted mean of our five measured HII regions, and shows the empirical relation of \cite{tremonti04}. The total effective yield is computed from the observed metallicity, $Z$, and the galaxy total gas fraction (not including dark matter, 0.70 for NGC 2915), $\mu$, such that y$_{eff}=Z/$ln$(\mu^{-1})$. In this case, 12 + Log (O/H), an oxygen abundance by number, is converted to a metallicity by mass using the conversion factor 11.728, assuming that helium accounts for 36\% of the total gas mass \citep{lee03}. The logarithm of the effective yield of NGC 2915 is  $-$2.516, given its error-weighted mean oxygen abundance of 12 + Log (O/H) =  7.97. Along with total stellar and gaseous masses presented in Table 1, NGC 2915 falls exactly where it is expected to fall on this plot. The relation of \cite{tremonti04} seen in Figure 4 as a solid line is attributed to metal loss via galactic winds. NGC 2915 lies near its turnover, where winds are thought to start playing an important role in blowing out metals.  In this context, NGC 2915 behaves just like other galaxies of its same mass, gas fraction, and metallicity.  
   
\begin{figure*}[t]
\epsscale{0.9}
\plotone{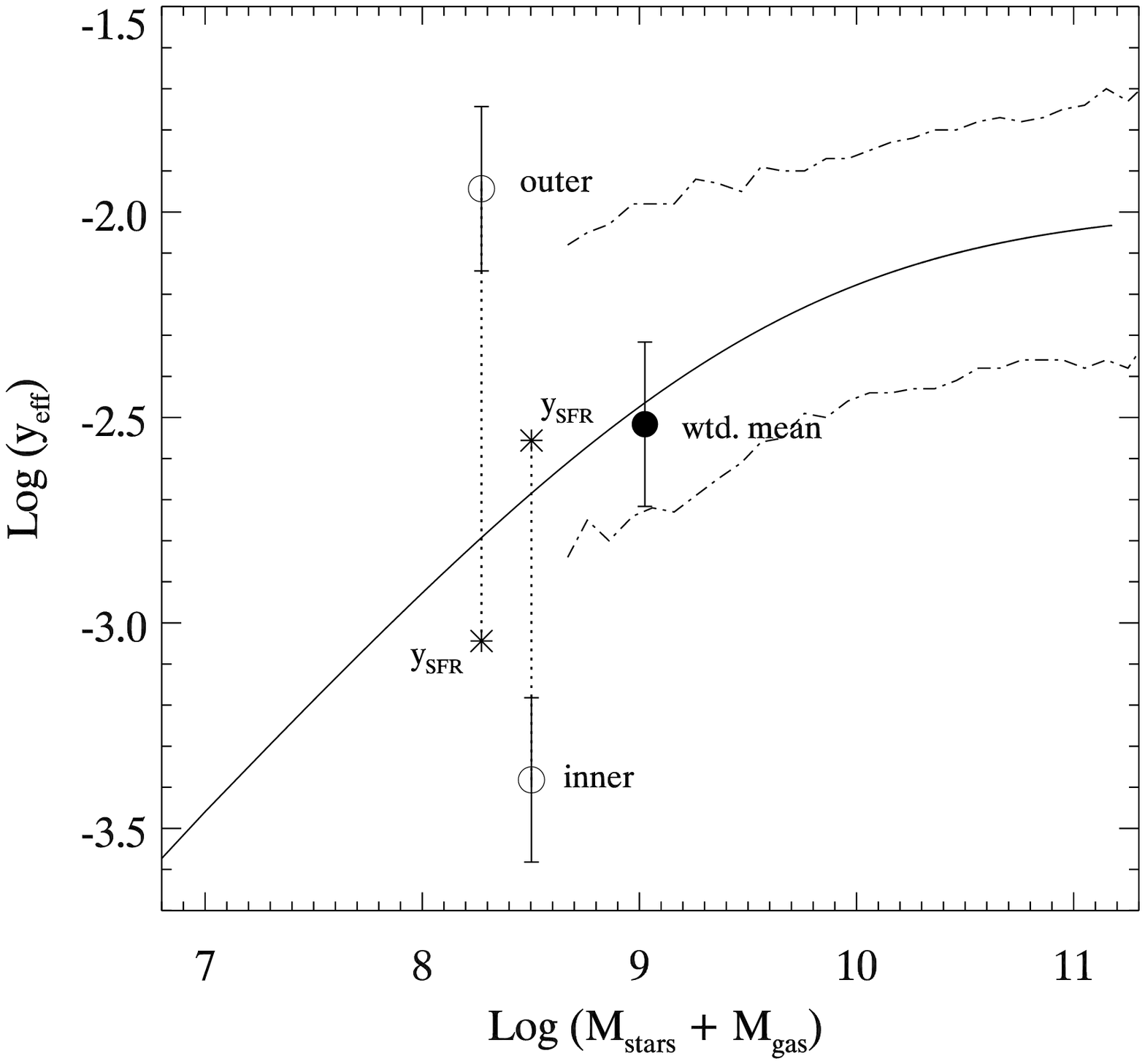}
\figcaption{\label{yields} The effective yield versus the total baryonic (stellar plus gas) mass for the mean 12 + Log (O/H) of NGC 2915 (filled circle), for the outer region of NGC 2915 defined in Section \ref{anal} (top open circle), and for the central, inner region of NGC 2915 as defined in Section \ref{anal} (bottom open circle). The empirical relation in \cite{tremonti04}, their equation 6, is shown for reference (solid line), along with contours (dashed-dotted line) that enclose 95\% of the SDSS data presented in \cite{tremonti04}. The vertical error bars (with hashes) for the points on this plot represent the error of 0.25 dex in the measured oxygen abundance, used here as a proxy for total metallicity. Assuming a 10$\%$ error in the stellar and gas mass estimates from M96, the radii of the NGC 2915 circles are the approximate size of the x-axis error. The dotted lines attached to each point extend to the effective yield values calculated from the current location-based FUV-derived SFRs (y$_{SFR}$) in Section \ref{anal}. These values show the extent to which NGC 2915's outer disk is overabundant for its current gas-fraction and SFR, and to which the inner-disk is similarly under-abundant. }
\end{figure*}
  Considering the effective yields of the inner and outer components of NGC 2915 separately, however, gives a very different result. We present two simple cases in which we calculate the oxygen abundance we might expect to measure in the outer gaseous disk of NGC 2915 based solely on its stellar and neutral gas content. In the first case, we estimate the effective yield in the outer-disk of NGC 2915, and compare it to the \cite{tremonti04} relation between effective yield and total mass. We assume the I-band light best traces the stellar mass, and calculate a radius centered on the galaxy that contains 90\% of the stellar mass (90\%\ of the I-band luminosity), r$_{90}$, to be $\sim44$". We then measure an approximate radius which contains all of the stellar light, including the outer HII regions and faint extended UV emission, r$_{tot}$, to be $\sim150$".   We note that ACS images presented in \cite{meurer03} show several globular clusters at radii similar to the outer HII regions. This population of older stars is included in our I-band measurements. 
  
  From the HI data presented in \cite{meurer96}, we find that r$_{90}$ contains 4.1\%\ of the total HI mass, and the annular region between r$_{90}$ and r$_{tot}$ contains 20.9\%\ of the total HI mass. Within this annular region, we measure an effective yield (log y$_{eff}$) of -1.94 and log (M$_{*}$ + M$_{gas}$) of 8.3. Based on the empirical relation of \cite{tremonti04}, we would expect log y$_{eff}$ $\sim$ -2.8 in the outer annulus. Based on these crude expectations, the outer regions of NGC 2915 are therefore overabundant for their gas fraction. The central region within  r$_{90}$ has log y$_{eff}$ $\sim$ -3.4, and  log (M$_{*}$ + M$_{gas}$) of 8.5, and is therefore concomitantly  under-abundant for its gas fraction. We have chosen not to include the entire HI disk in our outer annulus, and instead truncate it at the edge of the detectable I-band emission. Including the entire HI disk in this measurement would only make the discrepancies between the calculated and expected effective yields more extreme. 
  
  In the second case, we derive the current star formation rate density ($\Sigma_{SFR}$) of the region between r$_{90}$ and r$_{tot}$ from the GALEX far-UV flux in this region, assume a constant SFR and net oxygen yield over some star-formation timescale (see next paragraph), and use neutral gas surface densities from \cite{meurer96} to approximate an expected O/H ratio in the outer region. After subtracting foreground and background sources from the total flux, we measure the  FUV AB magnitude (uncorrected for extinction) within   r$_{90}$ to be 15.59$\pm0.06$ and that within r$_{tot}$ to be 15.48$\pm0.16$. The 1$\sigma$ errors here account explicitly for the rather large background non-uniformity in this low galactic latitude field ($b = -18.4^{\circ}$), and assume the outer disk contribution can never go negative.  To correct these values for extinction, we use the average of the E(B-V) values computed from the Balmer decrement (E(B-V)$_{avg}$ = 0.380), shown in Table 3, and the ratio of A($\lambda$)/E(B-V) for the FUV to be 8.24 \citep{wyder07}. Using the UV SFR conversion of \cite{salim07} for sub-solar metallicity, a Salpeter IMF, and star formation averaged over timescales of 100 Myrs, where SFR (M$_{\odot}$ yr$^{-1}$) = 1.08 $\times$ 10$^{-28}$ L$_{FUV}$, we find that $\Sigma_{SFR}$ = 3.4 $\times$ 10$^{-4}$ M$_{\odot}$ yr$^{-1}$ kpc$^{-2}$ in the outer annulus and $\Sigma_{SFR}$ = 3.5 $\times$ 10$^{-2}$ M$_{\odot}$ yr$^{-1}$ kpc$^{-2}$ in the central region. For reference, the values we derive for the entire disk of NGC 2915 are $\Sigma_{SFR}$ = 3.2 $\times$ 10$^{-2}$ M$_{\odot}$ yr$^{-1}$ kpc$^{-2}$, which amounts to a total SFR of 0.09 M$_{\odot}$ yr$^{-1}$, similar to the value derived by \cite{meurer94} of 0.05 M$_{\odot}$ yr$^{-1}$ using H$\alpha$ luminosities. 
  
  In order to be consistent with our measurements of the mean galaxy-wide oxygen abundance in NGC 2915 (12 + Log (O/H) = 7.97), we find the length of time over which to average the star formation to be 1.6 Gyr, assuming a net oxygen yield of 0.01 (Maeder 1992; \nocite{maeder92} the mass of oxygen ejected by all stars per unit mass of matter locked up in stars).  This timescale, along with an average gas density in NGC 2915 of 5.0 M$_{\odot}$ pc$^{-2}$ and $\Sigma_{SFR}$ given above, allows us to derive the galaxy-wide measured total oxygen abundance, 12 + Log (O/H), of 7.97. We then use this same timescale, along with measured SFRs, to calculate the oxygen generated in the outer annulus. Given $\Sigma$(gas) $\sim$ 3.2 M$_{\odot}$ pc$^{-2}$ at 100", a net oxygen yield of 0.01, and that star formation has been ongoing at the rate of $\Sigma_{SFR}$ = 3.4 $\times$ 10$^{-4}$ M$_{\odot}$ yr$^{-1}$ kpc$^{-2}$  for the last 1.6 Gyrs, the oxygen abundance in the outer annulus is expected to be 7.2 (0.04 Z$_{\odot}$). We have measured it to be 8.26 (0.4 Z$_{\odot}$), significantly higher than what is expected based on these crude assumptions. In order to reach our measured abundance in the outer disk of NGC 2915 at the current star formation rate, we would have to consider unrealistically-long timescales for ongoing star formation in the outer disk, longer than the estimated age of the universe. We also note that, under these same assumptions, with a $\Sigma$(gas) $\sim$ 8.5 M$_{\odot}$ pc$^{-2}$ at 25", the central region would be expected to have an oxygen abundance of 1.9 Z$_{\odot}$, 12 + Log (O/H) = 8.8, much higher than our measurement of 7.93 (0.2 Z$_{\odot}$).
  \section {Discussion}
  The flat, or possibly increasing oxygen abundance gradient in NGC 2915  leads us to conclude that  one or more physical process(es) has distributed metals throughout its extended gaseous disk.  The low-level of  current star formation observed at large radii in NGC 2915  is not sufficient to have produced the measured oxygen abundances. In order for the star formation to be responsible for the oxygen enrichment at large radii, it would have to have been going on for longer than the age of the universe.  \cite{bresolin09a} come to a similar conclusion regarding metal-mixing after finding no gradient in the extended UV disk of M83. However, they do note that if star formation in the outer disk has been ongoing for the last 2-3 Gyrs, then it could potentially explain the flat oxygen abundance gradient in the outer disk.   
 	
	The role of self-enrichment in the higher-than-expected oxygen abundances of the outer HII regions in NGC 2915 is most likely negligible. While the winds of massive stars can inject $\sim0.5$ M$_{\odot}$ of oxygen in {\emph{metal-enriched}} HII regions  over the course of their brief lives, their impact is minimal in HII regions with originally low metallicity (Z $<$ 0.4 Z$_{\odot}$; Meynet \& Maeder 2005\nocite{mm05}). The metal lines are required for driving the winds efficiently. Along these lines, Wofford (2009) explores the impact of massive star winds on low-metallicity (Z$\sim0.05$ Z$_{\odot}$), massive HII regions (10$^{6}$ M$_{\odot}$) using starburst99 and CLOUDY codes, and finds that they contribute a maximum abundance enhancement of $\Delta$Log (O/H) $\sim0.025$ dex. While the effect may be  larger for the low-luminosity outer HII regions presented in this work  due to their low gas mass and relative high fraction of O stars, it cannot account for the extent to which the outer HII regions are over-abundant. In this section, we consider 3 scenarios that may be responsible for the observed oxygen abundance trends in NGC 2915, and discuss the implications of our results. 
  
  \subsection{Scenario 1: Metal Mixing Within the HI disk}
 The lower-than-expected central oxygen abundances and the higher-than expected outer-disk oxygen abundances may support a metal mixing scenario, in which metals generated by the central star-forming core are transported outward though the disk. The chemodynamical model of \cite{ferguson01} shows that viscous flows, perhaps resulting from cloud-cloud collisions or gravitational instabilities, can transport metals to large radii, resulting in flat abundance gradients.  In simulations tailored to the Milky Way, \cite{minchev09} find that the interaction between spiral structure and a central bar is an effective and efficient (t $<$ 3 Gyrs)  mechanism for radial mixing in galactic disks, out to large radii. Yet, previous claims of a central massive HI bar in NGC 2915 have recently been discredited by the newer, better-resolution HI synthesis data of \cite{elson10}, who instead find two elongated central HI concentrations separated by only 1.1 kpc in the core. Nonetheless, there is still a bar-like structure in its central HI morphology, which may be partially responsible for directing gas flow outward, and producing a diluting effect and a shallow abundance gradient  \citep{friedli94,martinet97}. 


 \cite{tassis08} present a cosmological model that does not include supernovae-driven metal ejection, yet reproduces the mass-metallicity relation for simulated galaxies. Their primary physical explanation for the decreasing metallicity with decreasing galaxy mass is that star formation is increasingly inefficient in low mass systems (see also Dalcanton 2007\nocite{dalcanton07}). To explain the low effective yields of dwarf galaxies, they propose metal-mixing, and suggest the transport of metals from the inner regions of the disk to its outer, unobservable regions of the halo. Detections of these metals in the warm/hot gas of the halo would be available only through absorption line studies using background quasars (e.g. Tripp et al. 2008\nocite{tripp08}). The high effective yield in the outermost gaseous disk of NGC 2915 is consistent with the picture of metal transport provided by \cite{tassis08}, though does not confirm the presence of metals outside the HI-disk, in the ionized halo gas.

 \subsection{Scenario 2: Supernovae-Driven Blowout and Fallback}
 
It has long been noted that feedback from supernovae in the shallow potential wells of dwarfs may serve to efficiently blow out the metals generated during the stellar evolution process (e.g. Larson 1974; Dekel \& Silk 1986\nocite{larson74, dekel86}). Indeed, the low effective yields for dwarf galaxies are commonly attributed to strong galactic winds generated by supernovae (e.g. Tremonti et al. 2004\nocite{tremonti04}), and dwarf galaxies may therefore be at least partially responsible for the enrichment of the IGM \citep{stocke04}. In order for this metal blowout scenario to fit the observed radial oxygen abundance trends in NGC 2915, a large percentage of the metals would have to fall back at large radii, at least a few kpc from where they were ejected. In a sense, then, this scenario has the overall effect of the metal-mixing scenario outlined above, though it is powered by a completely different source.  Yet, most current models of supernovae-driven outflow tend to favor the ejection of metals into the IGM, rather than the fallback of metals on the galaxy's disk \citep{brooks07}. If these models do include the fallback of metals, this fallback generally occurs tens of parsecs from where the metals were originally ejected \citep{maclow99}. Further complicating this interpretation, results from \cite{summers03} indicate that a large HI halo may prevent the escape of a wind. 

  The lack of a model for a scenario in which metals are entrained in galactic winds, and fall back down at large galactocentric radii does not make such a scenario impossible. Some observational evidence presented in \cite{vcb05} for M82 suggests that such a scenario may indeed be possible. Moreover, our H$\alpha$ images of NGC 2915 do show a bubbly H$\alpha$ morphology (see Figure 1) that may be evidence for gas flows from the star-forming core.  Additionally, \cite{elson10} find high central velocity dispersions ($\sim30$ km/s) in new, more highly-resolved HI data from the ATCA  which could be indicating that the central gas dynamics of NGC 2915 are largely dominated by stellar winds from the central massive stars.  Wind velocities derived from X-ray temperatures in several dwarfs can range from 500$-$900 km/s \citep{martin02}. However, no such hot gas measurements exist specifically for NGC 2915.  The H$\alpha$ emission line widths of the central HII regions (regions 4, 5, and 6) do show some signs of line broadening, with a mean FWHM of 2.36 \AA~($\sim110$ km/s).  For reference, the mean H$\alpha$ FWHM of the outer HII regions is 1.9 \AA~($\sim85$ km/s), roughly consistent with the 2 $\times$ 2 binned spectral resolution of 1.7 \AA~($\sim75$ km/s at H$\alpha$). We note that an estimate of the escape velocity from the inner parts of NGC 2915, based on its rotation speed of 80 km/s derived by M96, where $v_{esc}\sim 3\times v_{circ}$ \citep{vcb05}, is 240 km/s. Based on these rough calculations, we see no convincing evidence in the H$\alpha$ line widths for the escape of ionized gas from the central starburst of NGC 2915.

 \subsection{Scenario 3: Past Interaction}
 
 NGC 2915 could have interacted with a low-surface-brightness neighbor and accreted its gas, potentially enriched from previous generations of star formation. In addition, the interaction may have triggered both the central star formation, and the faint outer-disk star formation in NGC 2915 \citep{hernquist95, hopkins09}. Although it is not always the case, outer-disk, extended star formation does tend to be associated with previous or ongoing interactions \citep{Thilker07,werk10}.  Nonetheless, even if the metal-enriched gas from the accreted galaxy explains the high oxygen abundances in the outer regions of NGC 2915, we would still need an additional mechanism (one of the above) to explain the relatively low oxygen abundance in the central, dominant star-forming region. Along this line, recent numerical simulations by \cite{rupke10} predict that oxygen abundances in the central parts of interacting galaxies will be lower than expected based on the mass-metallicity relation due to radial inflow of low-metallicity gas from the outskirts of the merging galaxy. In this model, subsequent radial mixing (see Scenario 1) tends to flatten the observed metallicity gradients. If NGC 2915 has undergone a merger in its distant past, it appears to fit the model of \cite{rupke10} very well. 
 
 
The relatively smooth kinematics and spiral structure of the extended HI disk requires that such an interaction happened at least 3 Gyr ago \citep{barnes96, hopkins09}.  M96 do find signs of a warp in the HI-disk, a bar, and dark-matter regulated (strong) star formation in NGC 2915, all of which could be attributed to an interaction with another galaxy (though such an interaction is not required). Although NGC 2915 is in a low density environment, M96 note that there is one possible interaction partner, the low-surface-brightness ``object" SGC 0938.1-7623 (KK98-076; Corwin et al. 1985, Karachentseva \& Karachentsev 1998\nocite{corwin85,kk98}). Based on observations with the 64-m Parkes Telescope, M96 note that SGC 0938.1-7623 is either not at the velocity of NGC 2915, or it is very gas poor. They give a 5$\sigma$ upper limit on M$_{HI}$ of the object to be 2.6$\times$ 10$^{6}$ M$_{\odot}$. Also noted in the original discovery paper \citep{corwin85} and in \cite{kk98} is the possibility that this object is a reflection nebula. While there is no explicit evidence of an interaction, we cannot rule it out. We additionally note that if the abundance is actually increasing with radius (see Section 4), our results would probably favor an accretion scenario in which metal-rich gas is preferentially deposited into the outer parts of the galaxy \citep{peek09} in conjunction with metals being transported out of the under-abundant central regions. 

  
  \subsection{Implications for the Origin of the Mass-Metallicity Relation}
  
 The three scenarios discussed above are not necessarily mutually exclusive. For instance, the metal mixing scenario (1) could additionally include the SN-driven ejection of metals from the central parts to the IGM (scenario 2), given the under-abundance of the central regions, and that the global properties of NGC 2915 fit well with previously-determined trends relating metallicity, effective yield, and mass that have been attributed to metal loss via strong galactic winds. Or, the interaction scenario (3) could additionally include radial metal mixing (scenario 1), as does the model of \cite{rupke10}. And finally, the interaction scenario (3), could include strong metal outflows from the central starburst (scenario 2), without the need for fallback.  
 
 Without further modeling the properties of the stars and gas in NGC 2915, we cannot say for certain which or what combination of these processes is contributing to the flat (possibly increasing) abundance gradient we observe. However, it is clear from the possibilities outlined that the radial redistribution of metals, whether driven by supernova-generated winds and fallback or viscous flows within the disk, is required to reproduce the flat abundance gradient.  Every scenario or combination of scenarios requires metal transport from the central to outer regions to some degree, with the exception of the interacting scenario (3) plus metal blow-out (2, without fallback) from the central regions. Our results, therefore, imply that metal-mixing is a significant process in the extended gas disk of NGC 2915. Furthermore, flat abundance gradients may not be uncommon in systems with low star formation efficiency and/or high HI content, perhaps owing to similar processes of metal-redistribution. In addition to the flat abundance gradient in the outer low surface brightness disk of M83 \citep{bresolin09a}, \cite{deblok98} find no oxygen abundance gradient in a sample of 12 low surface brightness galaxies over the radial range of 3 scale lengths.

 
   Our results indicate that a significant outflow of metals into the IGM may not be needed to reproduce the measured effective yields in NGC 2915.  We have presented two plausible scenarios that explain the abundances in NGC 2915 that do not include supernova-generated galactic winds. Ours is not the first work to cast doubt on the universality of dramatic metal loss in dwarf galaxies. For example,  a number of the dwarf galaxies in the sample of \cite{lee06} have yields far greater than expected from the empirical relation of \cite{tremonti04}, leading the authors to conclude that some less energetic form of mass loss may be at work, or that star formation efficiencies are low in dwarf irregulars.  Efficient metal mixing in extended gaseous disks is another physical mechanism that may be responsible for the observed low effective yields in dwarf galaxies. 
   
  
   \section{Summary and Conclusions}
   
   We have derived oxygen abundances from optical emission-line spectra of 5 HII regions in the extended gaseous disk of NGC 2915, in locations ranging from the central starburst to 1.2 times the Holmberg radius. Outer-disk HII regions were originally found using deep AAT H$\alpha$ images, and appear to lie in a very faint extended-UV disk visible in recent deep GALEX images. Out to these large galactocentric distances, we find no evidence of a decreasing metallicity gradient. The central HII regions have a metallicity of 0.2 Z$_{\odot}$  ($\pm$ 0.15 dex), while the outer HI regions appear to be enriched at a level of 0.4 Z$_{\odot}$  ($\pm$ 0.25 dex). Based on calculations of metal yields and star formation rates in NGC 2915, we conclude that the outer disk is considerably more metal-rich than its ``expected" value of 0.04 $-$ 0.08 Z$_{\odot}$, and that the central region is similarly under-abundant compared to its ``expected" value of 1.7 $-$ 1.9 Z$_{\odot}$. These observations indicate that some process, other than ongoing star formation, has enriched the gas at large radii. We present 3 plausible (and non-exclusive) scenarios for the metal-enriched outer gas disk of NGC 2915: metal mixing, supernovae-driven winds entraining metals and falling back down at large radii, or a past interaction leading to the accretion of enriched gas. Measurements of metal abundances for the outer gaseous disks of additional galaxies will cast light on the processes that  redistribute metals and their effect on the mass-metallicity relation for galaxies. 
   
   \section{Acknowledgements}
   
   JKW thanks Lisa Kewley, Liese van Zee, and Kevin Croxall for very helpful discussions of  strong-line oxygen abundance determinations and sources of error. Insights offered by Henry Lee, John Salzer, Sally Oey, and Rob Kennicutt helped to shape this work, and are much appreciated. An enthusiastic thank you to the Carnegie Scientists that maintain and support COSMOS (the IMACS multi-slit data reduction package), specifically Greg Walth, Dan Kelson, Alan Dressler, and Gus Oemler, for answering all questions about sky-subtraction, standard star observations, and installation/implementation issues promptly and thoroughly! JKW also thanks Mauricio Martinez, Magellan Telescope Operator, for being most efficient, keeping clouds away, and providing good company during long exposure times. Additionally, this paper benefited from helpful suggestions by the anonymous referee. MEP and JKW acknowledge support for this work through NSF CAREER AST-0904059, the Research Corporation, and support from the Luce Foundation.  GRM was partially supported by NASA LTSA grant NAG5-13083 for the work presented here.
   
   {{\it Facilities:} \facility{Magellan: Baade}, \facility{AAT},\facility{GALEX}, \facility{CTIO:1.5m}. }
\bibliography{references}
\bibliographystyle{apj}




\clearpage
\clearpage

\clearpage
\clearpage

\end{document}